\documentstyle[aps,epsf]{revtex}  
\def\desepsf(#1 width #2){\epsfxsize=#2 \epsfbox{#1}}
\def\greatersim{\ \hbox{\raise 2pt \hbox{$>$} \kern -13pt
                     \lower 3pt \hbox{$\sim$}}\ }
\begin{document}        
\baselineskip 14pt
\title{Hard diffraction from small-size color sources\footnote{Talk 
given by F.\ Hautmann at the Division 
of Particles and Fields Conference, Los Angeles, January 5-9, 1999.}} 
\author{F.\ Hautmann}
\address{Department of Physics, Pennsylvania State University, 
        University Park PA 16802, USA}
\author{Z.\ Kunszt}
\address{Institute of Theoretical Physics, 
        ETH, CH-8093 Zurich, Switzerland}
\author{D.E.\ Soper}
\address{Institute of Theoretical Science, 
        University of Oregon, Eugene OR 97403, USA}        
\address{and Theory Division, 
        CERN, CH-1211 Geneva 23, Switzerland}
%
\maketitle              

\vskip 0.6 cm 

\begin{abstract}        
     We describe diffractive hard processes in the framework of 
     QCD factorization and discuss what  one can  learn from the 
     study of hadronic systems with small transverse size.  
\
\end{abstract}   	

\vskip 1.2 cm

Diffractive deeply inelastic structure functions satisfy a 
factorization theorem of the form~\cite{proo,bere,fracture}
\begin{equation} 
\label{fact} 
F_2^{\rm diff} 
\sim  {\hat F}_{ a}  \otimes 
  { df_{a/A}^{\rm diff} 
\over dx_P\, dt} 
\hspace*{0.3 cm} ,
\end{equation}  
where the first factor on the right hand side is 
a short-distance scattering function and the second factor is 
a diffractive parton distribution, containing the long-distance physics. 
The short distance factor is no different  
than in inclusive deeply inelastic scattering. The long distance 
factor is.  
Although the evolution equation for the diffractive
parton distribution functions is the same as that of the inclusive
parton distribution functions~\cite{proo,bere,kust}, 
their behavior at a fixed scale $\mu_0$
that serves as the starting point for evolution may be very different
from the behavior of the inclusive functions. 
The different
phenomenology that characterizes diffractive versus inclusive deeply
inelastic scattering depends entirely on this.

Diffractive parton distributions in a proton  
at the scale $\mu_0$ are not perturbatively calculable. 
This is because the proton has a large transverse size. 
Suppose one had a
hadron of a size $1/M$ that is small compared to $1/\Lambda_{\rm {QCD}}$.  
Then one could compute diffractive parton distributions as a 
perturbation expansion. 
Results on diffraction of small-size hadronic systems have been presented 
in Ref.~\cite{hks}.

Fig.~1 shows a typical Feynman graph for one such case. 
Here we have considered a color-singlet current 
that couples only to heavy quarks of mass  $ M \gg 
\Lambda_{\rm {QCD}}$. This system gets diffracted and 
acts as a color source with small radius. 
This is represented in 
 the lower part of the graph.  
The top part of the graph represents the bilocal field 
operator~\cite{bere}  that defines the gluon distribution. 
The particular Feynman graph in Fig.~1, although of a rather high 
order in $\alpha_s$,   
is  leading in the limit 
$1 / x_P \to \infty$, where $x_P$ is the fractional  loss 
of longitudinal momentum from the diffracted hadron. 

The physical picture that emerges from the analysis of graphs like that
shown in Fig.~1 in the limit $1 / x_P \to \infty$ is that of the
familiar ``aligned jet'' model~\cite{align}. The bilocal operator
creates a large-momentum parton together with a color source of the
opposite color. This is confined to move on a lightlike line and is part of
the definition of the (inclusive or diffractive) parton distribution
functions. This happens far from the incoming small-size 
hadron. The system created by the  operator 
then passes through the color field of the small-size 
hadron, absorbing two gluons. What we have, then, is the scattering of
two color dipoles.

At this order of perturbation theory, 
the  result for 
the diffractive parton distributions  
 has the following form  

\begin{figure}[htb]
\centerline{ \desepsf(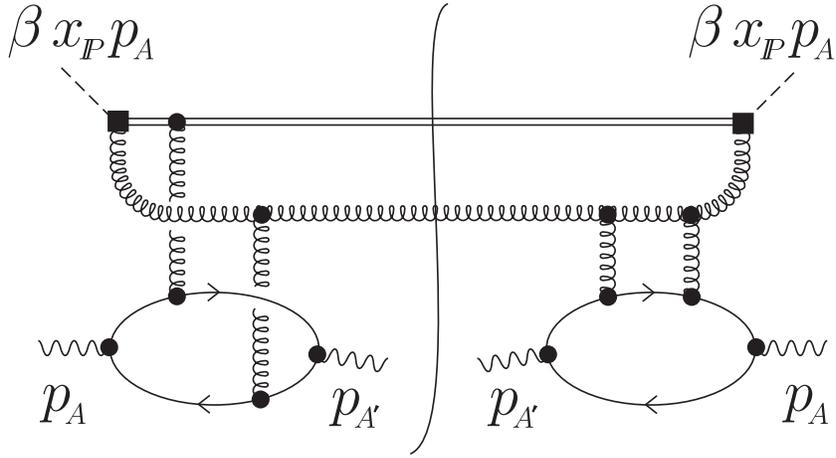 width 11 cm) } 
\caption{ A typical Feynman graph contributing to the 
diffractive gluon distribution of the model vector meson.  }
\label{figdiffpho}
\end{figure}

\vspace{10mm} 
 
\begin{equation}
\label{hresc}
{{d f_{a/A}^{\rm {diff}} 
 } \over
{dx_P \,dt}} (\beta ,  x_P , {\mbox{\bf q}}^2 , M
)
 = 
{ { \alpha^2 \, e_Q^4 \, \alpha_s^4  } \over { x_P^2 \, M^2 }  }\, 
h_a (\beta , {\mbox{\bf q}}^2 / M^2)  \, \left[ 1 + {\cal O} (x_P) \right]
\hspace*{0.2 cm} .   
\end{equation} 
Here  $ \beta x_P$ is the fraction of 
the hadron's longitudinal momentum carried by the parton and  ${\bf q}$ 
 is the diffracted transverse momentum ($ t \approx - {\bf q}^2   $).  
  The functions $h_a$ are 
plotted in Fig.~2.  

\begin{figure}[htb]
\vspace{115mm}
\includegraphics{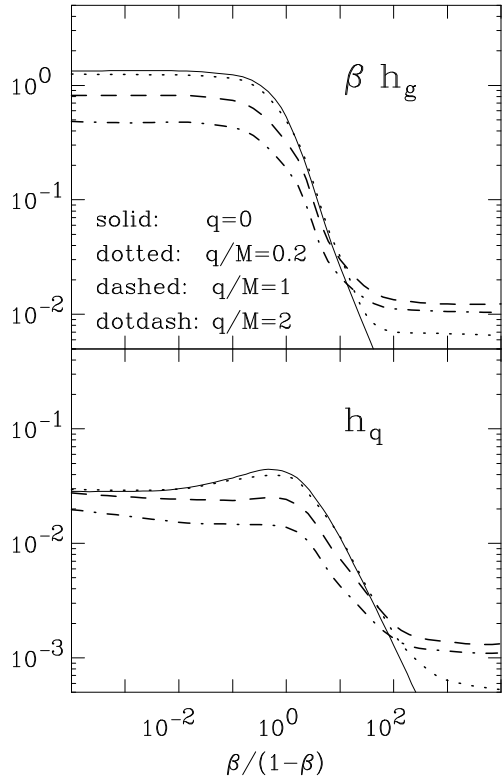}
\caption{ The $\beta$ dependence 
of the  gluon (above) and  quark (below) diffractive 
distributions for different values of 
${\mbox{\bf q}}^2 \simeq | t |$. 
 The rescaled distributions $h_a$ are defined 
 in Eq.~(2). }
\label{fighlogbeta}
\end{figure}

For small $\beta$ the distributions behave as 
\begin{equation}
\label{hb0}
h_g \propto \beta^{-1} \hspace*{0.1 cm} ,  
\hspace*{0.3 cm} h_q \propto \beta^{0} \hspace*{0.5 cm} 
(\beta \to 0) \hspace*{0.2 cm} . 
\end{equation} 
For large $\beta$, 
both  the   
gluon and quark 
 distributions evaluated at any finite ${\mbox{\bf q}}$ 
 have a constant behavior: 
\begin{equation}
\label{hb1} 
  h_g ,  h_q 
 \propto (1 - \beta)^0 \hspace*{0.5 cm} 
(\beta \to 1 , \hspace*{0.1 cm} 
 {\mbox{\bf q}} \neq 0 ) \hspace*{0.2 cm} . 
\end{equation} 
 At ${\mbox{\bf q}} = 0$ there are 
cancellations in the 
leading $\beta \to 1$ coefficients, so that the distributions 
vanish in the $\beta \to 1$ limit:  
\begin{equation}
\label{hb1q0}
h_g \propto (1-\beta)^{2} \hspace*{0.1 cm} ,  
\hspace*{0.3 cm} h_q \propto (1-\beta)^{1} \hspace*{0.5 cm} 
(\beta \to 1 , \hspace*{0.1 cm} 
 {\mbox{\bf q}} = 0 ) \hspace*{0.2 cm} . 
\end{equation} 
 The distributions are dominated by small 
$ | t | \simeq  {\mbox{\bf q}}^2 $ everywhere except 
at very large $\beta$.
Note that 
\begin{equation}
\label{gmuchbgr}
h_g \gg h_q 
\hspace*{0.2 cm} . 
\end{equation} 
Roughly, the order of magnitude of the ratio 
between the diffractive gluon and quark distributions  
can be accounted for 
by the ratio of the 
associated color factors, $ C_A^2 \, (N_c^2 -1) /[ C_F^2 \, N_c ] = 27/2$.

The result given above does not describe scaling violation. 
When additional gluons are emitted from the top subgraph in Fig.~1, 
 on the other hand, 
ultraviolet divergences  arise. The renormalization of these divergences 
leads to the dependence of the diffractive parton distributions on a  
renormalization scale $\mu$. This dependence is  
governed by the renormalization group evolution equations. 
The higher order, ultraviolet divergent graphs are suppressed compared to
the graphs of the type in Fig.~1 
 by a factor $\alpha_s\,\log(\mu^2/M^2)$.
When $\log(\mu^2/M^2)$ is large, these contributions are important, and thus
evolution is important. On the other hand, when $\mu$ is of the same order as
the heavy quark mass $M$, the higher order contributions are small corrections 
to the graphs considered above. Thus one may interpret the result given above 
as a result for the diffractive  parton distributions at a
fixed scale of order $\mu^2 \approx M^2$.  Then the diffractive parton
distributions at higher values of $\mu^2$ are given by solving the evolution
equations with the result of Eq.~(2) as a boundary condition.

How are these calculations for small-size  systems related to the 
real world? Obviously, the protons probed in 
deeply inelastic scattering experiments at HERA 
have a large transverse size. 
Suppose that one had available a hadron of adjustable size. 
Start with a very small size,   
in which case diffraction is forced to take place mostly on short distances,  
and let the size increase. 
Since this scale acts as a physical infrared cut-off, 
longer and longer distances are now allowed to contribute to the 
diffraction process.  
In a naive perturbation expansion, by the time one gets to
 $1 \; {\mbox {fm}}$ (about the proton radius) the answer  
would be completely dominated by the soft region.  
On the other hand, as the size of the hadronic system 
increases,  nonperturbative dynamics sets in. 
The infrared-sensitive behavior suggested by the perturbative 
power counting is likely smoothed out by this dynamics. 
We  hypothesize that,   
as we go to larger and larger sizes, 
the distance 
scales that dominate the diffraction process, rather than 
continuing to grow,  
stay frozen at some intermediate, semihard scale.  
The effect of this is to enhance the contribution from hard physics 
with respect to the contribution from soft physics. 

Note that recent experimental observations  
 on the $x_P$ dependence in 
diffraction may be regarded as providing support for the hypothesis 
of dominance of semihard scales in the diffractive parton distributions. 
It has been stressed~\cite{muedis98}  that the value of 
$\alpha_P (0) - 1$  (where $\alpha_P (0)$ is the pomeron intercept) 
measured in diffractive deeply inelastic 
scattering~\cite{h1scalviol,zeusf2d2} differs by a factor of $2$ 
from the corresponding value measured in soft hadron-hadron cross sections.

We may explore this hypothesis by investigating whether  
the short-distance result that we have found 
for small-size systems is also relevant to describe  
the physics of diffraction from large-size objects, such as protons at HERA. 
In particular, we are interested in two features of the HERA data for 
$F_2^{\rm diff} $~\cite{h1scalviol,zeusf2d2}: 
the surprising delay in the fall-off with $Q^2$ 
 and the surprising flatness in $\beta$.

To carry out this  study, 
we set $M$ in Eq.~(2) equal to 
$1.5 \, {\mbox {GeV}}$   and 
take the scale dependence of the diffractive parton distributions to
be that given by the two-loop evolution equations 
with the results (2) as a boundary condition at $\mu = M$.   
The choice of the value for $M$ corresponds to choosing 
a value for the semihard scale discussed above. 
The value of this scale, strictly speaking, is to be regarded as a 
free parameter  to be adjusted phenomenologically. 
In setting the  value to $1.5 \, {\mbox {GeV}}$  
we are guided 
by the expectation that this scale should be roughly of the order of a 
${\mbox {GeV}}$. See below for some qualitative remarks on different 
choices.   

The upper panel of Fig.~3 shows the result we obtain for 
the scale dependence of the diffractive 
(flavor-singlet) quark distribution at a fixed value of $\beta$, 
$\beta = 0.2$. 
(In this figure the dependence 
on $   {\mbox{\bf q}}^2 \simeq | t |$ is integrated over,   
from $0$ to $M^2$). By comparison, in the lower panel we 
show the analogous result for the ordinary 
(inclusive) quark distribution, taken from the standard set 
CTEQ4M~\cite{cteq4}.  
Fig.~3 illustrates the different pattern of scaling violation 
in diffractive and inclusive deeply inelastic scattering. 
At moderate values of momentum fractions, 
 while  the ordinary quark distribution is flat or 
 weakly decreasing with $Q^2$, the 
diffractive 
 distribution is rising with $Q^2$. 
The explanation for the rise 
 in the diffractive case 
lies with the behavior of the 
distributions at the initial scale $ M$
(Fig.~2). More precisely, it depends on the 
 gluon distribution being dominant throughout the range of 
momentum fractions.  
As $Q^2$ increases, 
gluons splitting into $  q {\bar q}$ pairs feed 
the quark distribution and cause it to grow  
in the region of moderately large  $\beta$.

\begin{figure}[htb]
\vspace{115mm}
\includegraphics{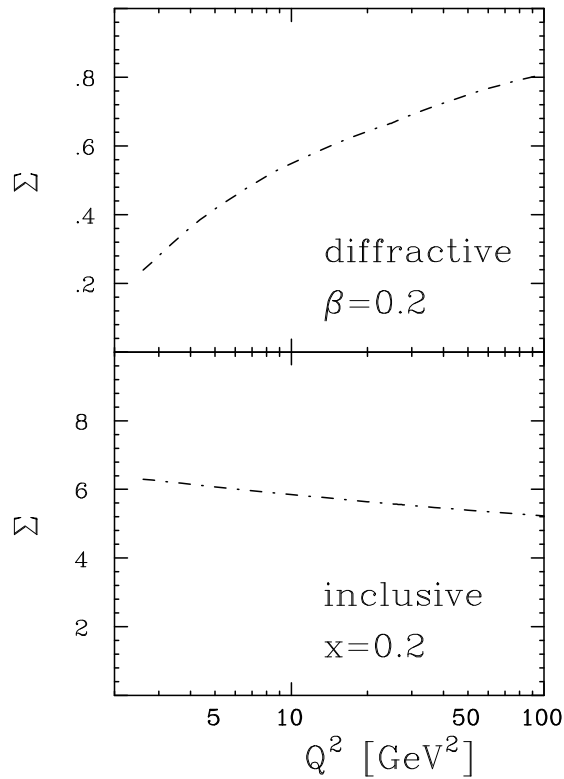}
\caption{ Scaling violation in the (flavor-singlet) quark 
distribution $\Sigma$ at moderate values of momentum fractions.  
Above is the case of the  diffractive  distribution, below is the 
case  of the inclusive distribution (from the set CTEQ4M).  }
\label{figcompqdep}
\end{figure}

In Fig.~4 we show results for the diffractive structure function 
 $F_2^{\rm diff}$ 
as functions of $\beta$ at various values of $Q^2$.    
In Fig.~5 we show the same results along with 
the ZEUS data~\cite{zeusf2d2}.  
Notice the main qualitative features of these results. 
 The sign of the scaling violation is positive up to 
 about $\beta \approx 0.55$, 
 reflecting the behavior noted for 
the quark distribution $\Sigma$ in Fig.~3.  
 In the range of intermediate values 
of $\beta$ (centered about $\beta \approx 0.5$) 
$F_2^{\rm diff}$  is rather flat in $\beta$. 
These features are distinctive of the diffractive structure function 
compared to the inclusive structure function. 

These features are in qualitative agreement with what is seen in the 
HERA data (Fig.~5). Note that, once one combines 
the analysis of diffraction from small-size states 
 with 
the hypothesis of dominance of 
semihard scales (with a particular scale choice),  
 all the dependences of the structure function 
 are fully determined.  
 That is,  not only the $ Q^2 $ dependence 
 but also the $\beta $ dependence are determined from theory. 
 (The same applies to the $t$ dependence. In 
\begin{figure}[htb]
\vspace{95mm}
\includegraphics{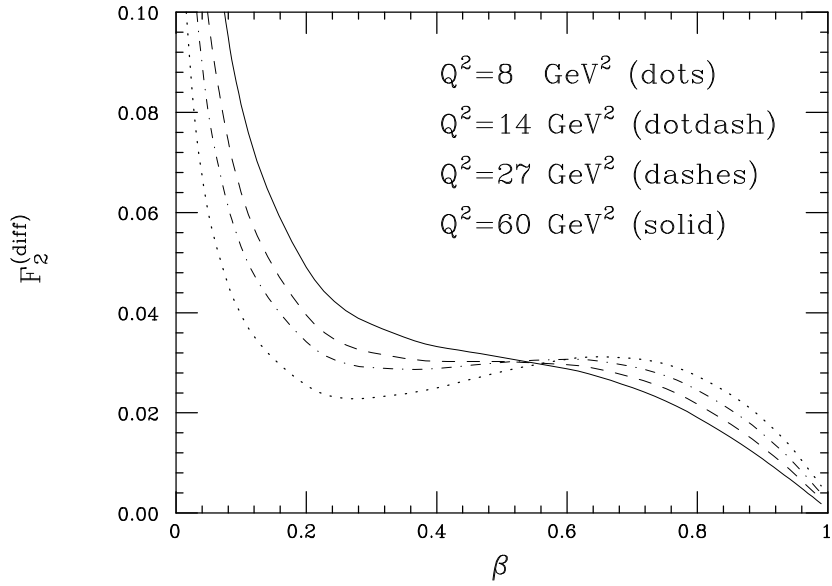} 
\caption{ The $\beta$ dependence 
of the  diffractive structure function $F_2^{\rm {diff}}$ 
for different values of $Q^2$. We compute $F_2^{\rm {diff}}$ 
in next-to-leading order.   } 
\label{figevf2nl}
\end{figure} 
\begin{figure}[htb]
\vspace{105mm}
\includegraphics{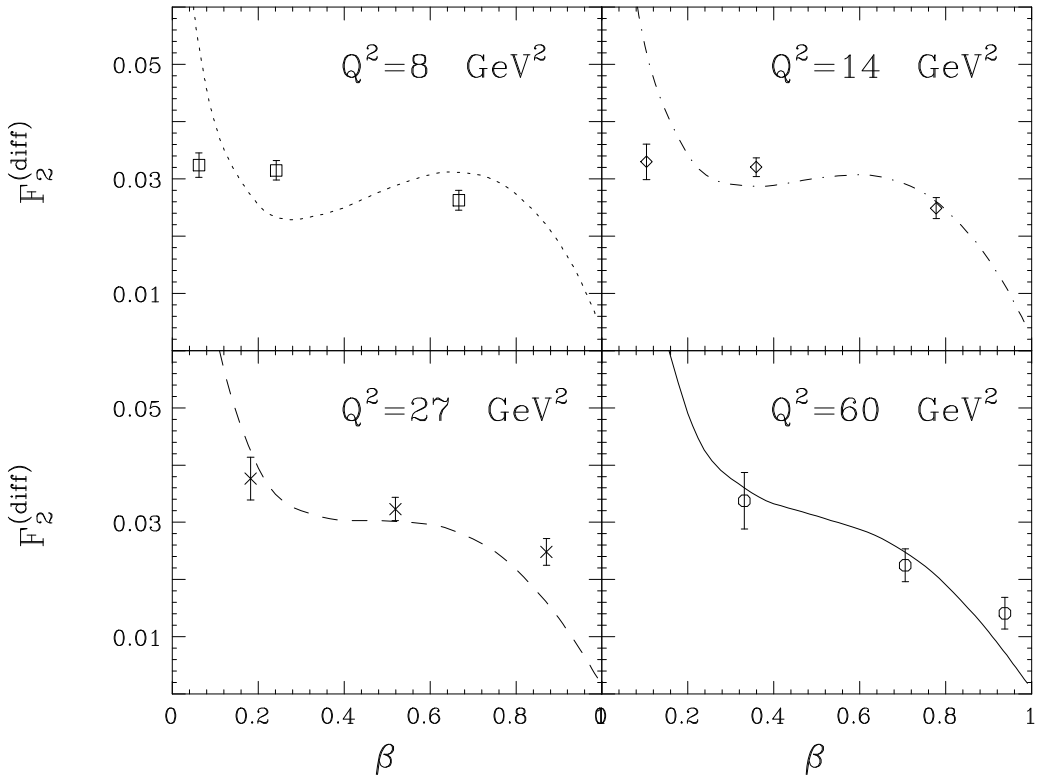}
\caption{ Same as in Fig.~4. Also shown are the ZEUS data 
from Ref.~[9]. }
\label{figfourpanel}
\end{figure} 
{\noindent the} results presented here 
 $t$ is integrated over.)  The only  free  
parameter left 
is an overall normalization,  
which has been adjusted arbitrarily in Figs.~4,5.

In the region of small $\beta$,  the curves of 
Figs.~\ref{figevf2nl},\ref{figfourpanel}  
have a different behavior from that suggested by the 
two data points at the lowest values of $\beta$ and lowest values of 
$ Q^2 $ ($ Q^2 = 8 \, {\mbox {GeV}}^2$ and $ Q^2 = 14 \, {\mbox {GeV}}^2$). 
If further data were to  confirm this difference, this 
could point to  interesting effects.     
Here we limit ourselves to a few qualitative remarks. 
As far 
 as the theoretical curves are concerned, 
we note that the 
diffractive distributions that  
serve as a starting point for 
the evolution are  fairly mild  as $\beta \to 0$. The gluon 
distribution goes like $1 / \beta$, while the quark distribution 
goes like a constant (see Eq.~(\ref{hb0})). The 
small-$\beta$ rise 
of the 
structure function $F_2^{\rm diff}$ 
in the curves of Figs.~\ref{figevf2nl},\ref{figfourpanel}    
is essentially due 
   to the 
form of the perturbative evolution kernels.  
As regards the data, it has been 
observed~\cite{ellisross} that 
for small $\beta$ the experimental identification of 
the rapidity gap signal may be complicated 
by the presence of low $p_\perp$ particles in the 
final state.  
If the current data hold up and especially if the same features are  
observed at lower values of $\beta$, it would  be interesting to see 
whether detailed models for the 
saturation of the unitarity bound~\cite{satura} 
could accommodate this small $\beta$ behavior. 

It is of interest to study how the comparison  
in  Fig.~\ref{figfourpanel} changes as the value of the 
semihard 
scale $M$ is 
changed. For example, one may be interested  to 
 lower this scale with respect to the value 
of $1.5 \, {\mbox {GeV}}$ used so far. 
For $M = 1 \, {\mbox {GeV}}$ we find  
 that, roughly speaking, the overall description of 
the data is of comparable quality (except at 
the lowest $\beta$ 
 and  $Q^2$ 
values, where 
the discrepancy noted above becomes more pronounced). 
However, we also find,  
in particular from the 
 modest steepness of the $\beta$ shape at the highest $Q^2$,  that 
a scale a bit higher than $1 \, {\mbox {GeV}}$ seems to be preferred, 
perhaps suggesting that the pomeron is a relatively ``small'' object. 
We do not push this study to a quantitative level at present, but 
it appears that in the future    more and better data on 
 diffraction  could  
be able to  give us useful information on the value of this scale. 
It would  be very interesting if one could connect  
it to other nonperturbative scales that enter in related areas of hadronic 
physics. 

\vskip 0.6 cm 

This research is supported 
in part by the U.S. Department of Energy grants No. DE-FG02-90ER40577 
and No. DE-FG03-96ER40969.

\end{document}